\documentclass{iaus}
\usepackage{graphics}

  \checkfont{eurm10}
  \iffontfound
    \IfFileExists{upmath.sty}
      {\typeout{^^JFound AMS Euler Roman fonts on the system,
                   using the 'upmath' package.^^J}%
       \usepackage{upmath}}
      {\typeout{^^JFound AMS Euler Roman fonts on the system, but you
                   dont seem to have the}%
       \typeout{'upmath' package installed. iaus.cls can take advantage
                 of these fonts,^^Jif you use 'upmath' package.^^J}%
      }
  \else
  \fi

  \checkfont{msam10}
  \iffontfound
    \IfFileExists{amssymb.sty}
      {\typeout{^^JFound AMS Symbol fonts on the system, using the
                'amssymb' package.^^J}%
       \usepackage{amssymb}%
         
         \let\geq=\geqslant
      }{}
  \fi

  \IfFileExists{amsbsy.sty}
    {\typeout{^^JFound the 'amsbsy' package on the system, using it.^^J}%
     \usepackage{amsbsy}}
    {}

  \usepackage{psfig}
  \usepackage{epsfig}

\newsavebox{\astrutbox}
\sbox{\astrutbox}{\rule[-5pt]{0pt}{20pt}}

\newcommand \Mpc {h^{-1}{\rm Mpc}}

\newcommand \kms {{\rm km~s}^{-1}}

\newcommand \mlsun {h M_\odot/L_\odot}
\newcommand \beqn {\begin{equation}}
\newcommand \eeqn {\end{equation}}

\title[Outskirts of Galaxy Clusters: intense life in the suburbs]
      {The Distribution of Mass and Light in Cluster Infall Regions}

\author[K. Rines]%
{K. Rines}

\affiliation{Yale Center for Astronomy and Astrophysics, Yale University,
New Haven, CT 06511, USA email: krines@astro.yale.edu \\[\affilskip]
}

\pubyear{2004}
\volume{195}
\pagerange{1--7}
\date{?? and in revised form ??}
\jname{Outskirts of Galaxy Clusters: intense life in the suburbs}
\editors{A. Diaferio, ed.}
\begin{document}

\maketitle

\begin{abstract}
The CAIRNS (Cluster And Infall Region Nearby Survey) project is a
large spectroscopic survey of the infall regions surrounding nine
nearby rich clusters of galaxies.  I describe the survey and use the
kinematics of galaxies in the infall regions to estimate the cluster
mass profiles.  At small radii, these mass profiles are consistent
with independent mass estimates from X-ray observations and Jeans
analysis.  I demonstrate the dependence of mass-to-light ratios on
environment by combining these mass profiles with Two-Micron All-Sky
Survey (2MASS) photometry.  Near-infrared light is more extended than
mass in these clusters, suggesting that dense cluster cores are less
efficient at forming galaxies and/or more efficient at disrupting
them.  At large radii, galaxy populations in cluster infall regions
closely resemble those in the field.  The mass-to-light ratio at these
radii should therefore be a good probe of the global mass-to-light
ratio.  The mass-to-light ratio in the infall region yields a
surprisingly low estimate of $\Omega_m \sim 0.1$.
\end{abstract}

\firstsection 
\section{Introduction}
The relative distribution of matter and light in the universe is one
of the outstanding problems in astrophysics.  Clusters of galaxies,
the largest gravitationally relaxed objects in the universe, are
important probes of the distribution of mass and
light. \cite{zwicky1933} first computed the mass-to-light ratio
of the Coma cluster using the virial theorem and found that dark
matter dominates the cluster mass.  Recent determinations using the
virial theorem yield mass-to-light ratios of $M/L_{B_j}\sim 250
\mlsun$ (Girardi et al.~2000 and references therein).  Equating the
mass-to-light ratio in clusters to the global value provides an
estimate of the mass density of the universe; this
estimate is subject to significant systematic error introduced by
differences in galaxy populations between cluster cores and lower
density regions (\cite{cye97,g2000}).  Indeed, some numerical
simulations suggest that cluster mass-to-light ratios exceed the
universal value (\cite{diaferio1999,kk1999,bahcall2000}).

Determining the global matter density from cluster mass-to-light
ratios therefore requires knowledge of the dependence of mass-to-light
ratios on environment.  Bahcall et al.~(1995) show that mass-to-light ratios
increase with scale from galaxies to groups to clusters.  Ellipticals
have larger overall values of $M/L_B$ than spirals, presumably a
result of younger, bluer stellar populations in spirals.
At the scale of cluster virial radii, mass-to-light ratios appear to
reach a maximum value.  Some estimates of the mass-to-light ratio on
very large scales ($>$10$\Mpc$) are available (\cite{bld95}), but the systematic uncertainties are large.

There are few estimates of mass-to-light ratios on scales between
cluster virial radii and scales of 10$\Mpc$
(Rines et al.~2000, Rines et al.~2001a, Biviano \& Girardi 2003,
Katgert et al.~2003, Kneib et al.~2003, Rines et al.~2004, Tully 2004
and references therein).  On these scales, many galaxies near clusters
are bound to the cluster but not yet in equilibrium (\cite{gunngott}).
These cluster infall regions have received relatively little scrutiny
because they are mildly nonlinear, making their properties very
difficult to predict analytically.  However, these scales are exactly
the ones in which galaxy properties change dramatically (e.g.,
Ellingson et al.~2001, Lewis et al.~2002, Gomez et al.~2003, Treu et al.~2003, Balogh et al.~2004, Gray et al.~2004).
Variations in the mass-to-light ratio with environment could have
important physical implications; they could be produced either by a
varying dark matter fraction or by variations in the efficiency of
star formation with environment.  In blue light, however, higher star
formation rates in field galaxies could produce lower mass-to-light
ratios outside cluster cores resulting only from the different
contributions of young and old stars to the total luminosity
(\cite{bahcall2000,tully}).

          Galaxies in cluster infall regions produce sharp features in
redshift surveys.  Early investigations of this infall pattern focused
on its use as a direct indicator of the global matter density
$\Omega_m$.  Unfortunately, random motions caused by galaxy-galaxy
interactions and substructure within the infall region smear out this
cosmological signal (Diaferio \& Geller 1997, Vedel \& Hartwick 1998).
Instead of sharp peaks in redshift space, infall regions around real
clusters typically display a well-defined envelope in redshift space
which is significantly denser than the surrounding environment
(\cite{cairnsi}, hereafter Paper I, and references therein).
 
Diaferio \& Geller (1997) and Diaferio (1999) analyzed the dynamics of infall regions with numerical simulations
and found that in the outskirts of clusters, random motions due to
substructure and non-radial motions make a substantial contribution to
the amplitude of the caustics which delineate the infall regions. Diaferio \& Geller (1997)
showed that the amplitude of the caustics is a measure of the escape
velocity from the cluster; identification of the caustics therefore
allows a determination of the mass profile of the cluster on scales
$\lesssim 10\Mpc$.

Diaferio \& Geller (1997) and Diaferio (1999) show that nonparametric
measurements of caustics yield cluster mass profiles accurate to
$\sim$50\% on scales of up to 10 $h^{-1}$ Mpc.  This method assumes
only that galaxies trace the velocity field. Indeed, simulations
suggest that little or no velocity bias exists on linear and mildly
non-linear scales (\cite{kauffmann1999a,kauffmann1999b}).  The caustic
method has been applied to systems as large as the Shapley
Supercluster (\cite{rqcm}) and as small as the Fornax cluster
(\cite{drink}) as well as to many nearby clusters (Paper I).  Biviano
\& Girardi (2003) applied the caustic technique to an ensemble cluster
created by stacking redshifts around 43 clusters from the 2dF Galaxy
Redshift Survey.  Rines et al.~(2000) found an enclosed mass-to-light
ratio of $M/L_R \sim 300 h$ within 4$~\Mpc$ of A576.  Rines et
al.~(2001) used 2MASS photometry and the mass profile from Geller et
al.~(1999) to compute the mass-to-light profile of Coma in the K-band.
They found a roughly flat profile with a possible decrease in $M/L_K$
with radius by no more than a factor of 3.  Biviano \& Girardi (2003)
find a decreasing ratio of mass density to total galaxy number
density.  For early-type galaxies only, the number density profile is
consistent with a constant mass-to-light (actually mass-to-number)
ratio.

Here, we calculate the infrared mass-to-light profile within the
turnaround radius for the CAIRNS clusters (Paper I), a sample of nine
nearby rich, X-ray luminous clusters.  We use photometry from 2MASS,
the Two Micron All Sky Survey (\cite{twomass}) and add several new
redshifts to obtain complete or nearly complete surveys of galaxies up
to 1-2 magnitudes fainter than $M^*_{K_s}$ (as determined by Cole et
al.~2001 and Kochanek et al.~2001).  Infrared light is a better tracer
of stellar mass than optical light; it is relatively insensitive to
dust extinction and recent star formation.  Despite these advantages,
there are very few measurements of infrared mass-to-light ratios in
clusters (\cite{tustin,rines01a,lin03,cairnsii}).

\begin{figure}
\centerline{\psfig{file=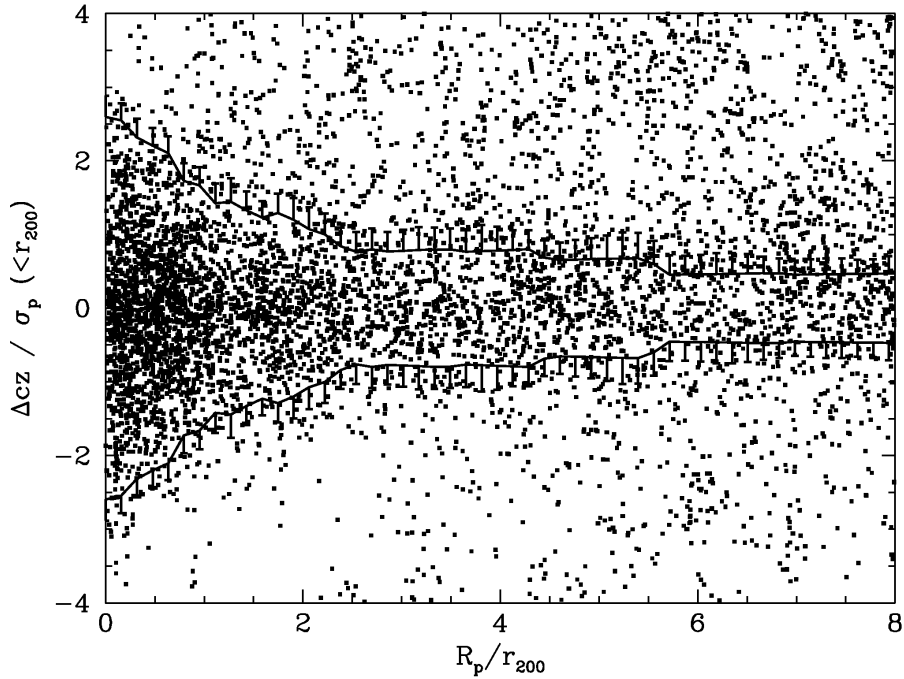,angle=0,width=12.5cm}}
  \caption{Redshift versus projected radius for the combined CAIRNS cluster.  The solid lines indicate the caustics and the errorbars show 1-$\sigma$ uncertainties.}
\label{fig:caustics}
\end{figure}

\section{Observations}\label{sec:observations}

The CAIRNS clusters are 8 of the 14 Abell clusters that are: nearby
($cz<15,000~\kms$), Abell richness class $R\geq1$, X-ray luminous
($L_X>2.5 \times 10^{43} h^{-2}$erg s$^{-1}$) galaxy clusters with
declination $\delta>-15^\circ$.  Between 1997 and 2003, we collected
5607 redshifts (both new and remeasured) in large areas around nearby
clusters with the FAST spectrograph on the FLWO 1.5-m telescope in
Arizona.  We targeted galaxies within a projected radius of
$\sim$10$\Mpc$ of the clusters, selecting targets first from POSSII
103aE plates and later from 2MASS when it became available.  The
redshift catalogs in Rines et al.~(2004) complete $K_s$ band
magnitude-limited samples extending 1-2 magnitudes fainter than the
characteristic magnitude $M_{K_s}^*$.  These samples include
$\sim$60-85\% of the total light in the clusters and their infall
regions.

\section{Mass Profiles}\label{sec:massprof}

Figure \ref{fig:caustics} shows the redshifts of galaxies surrounding
nearby clusters as a function of projected radius (normalized to
$r_{200}$, the radius within which the average density is 200 times
the critical density).  The caustic pattern (a dense envelope in phase
space with well-defined edges) is evident in the combined cluster as
well as in each of the 9 clusters.  Using the phase space distribution
of galaxies in cluster infall regions, we apply the kinematic mass
estimator of D99 to these clusters.  The resulting mass profiles agree
well with NFW (Navarro et al.~1997) and Hernquist (1990) models, but
exclude an isothermal sphere.

These mass profiles agree with X-ray masses at small radius as well as
virial masses (after correction for the surface pressure term) at
slightly larger radii (\cite{rines02}, Paper I).  The latter is
primarily a consistency check as the caustic technique utilizes the
same kinematic data as the virial theorem.  This consistency can be
further demonstrated with the velocity dispersion profiles (see Paper
I and Rines et al.~2004).

\section{Near-Infrared Luminosity Functions}\label{sec:klf}

\begin{figure}
\centerline{\psfig{file=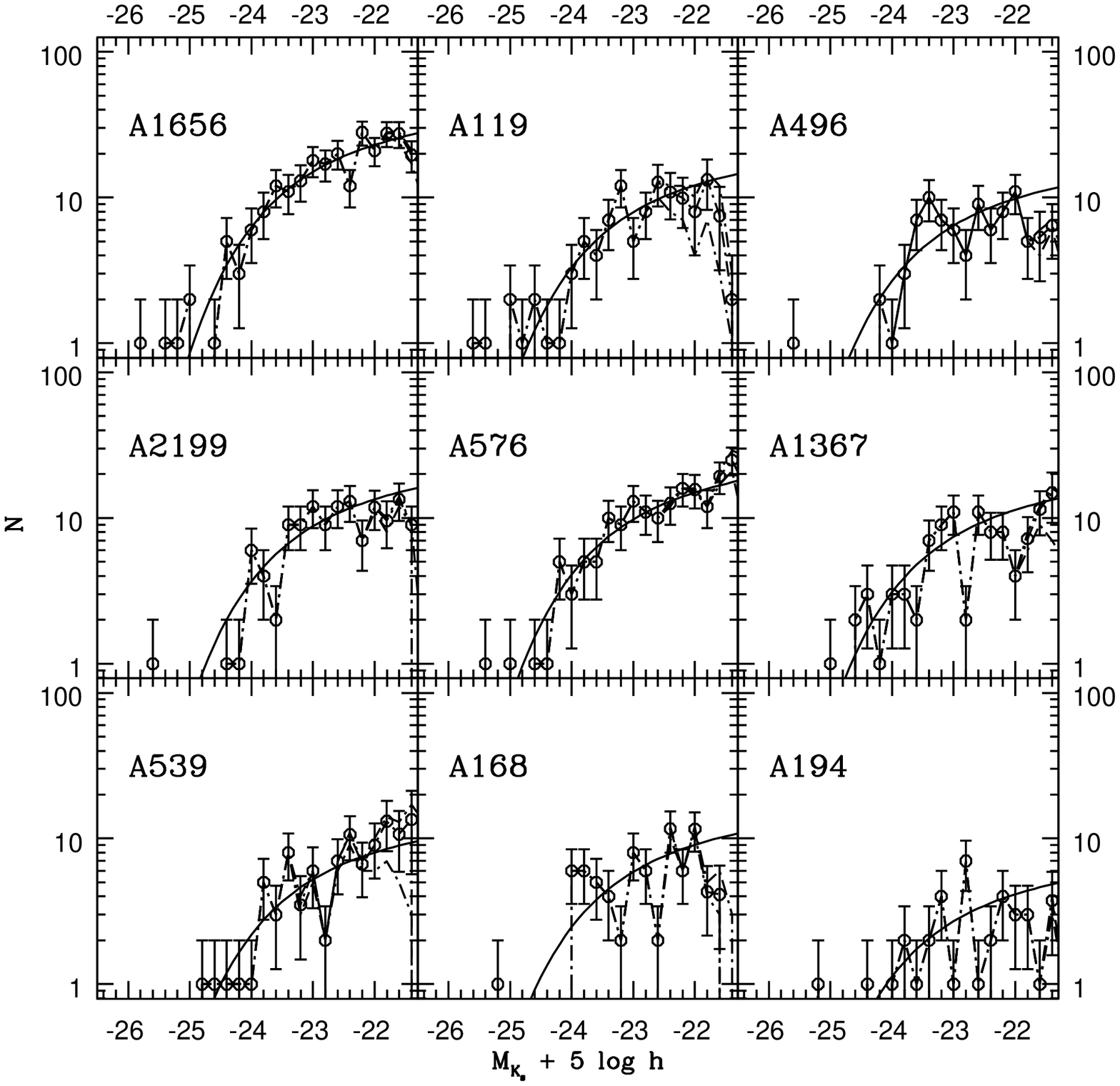,angle=0,width=6.5cm}
\psfig{file=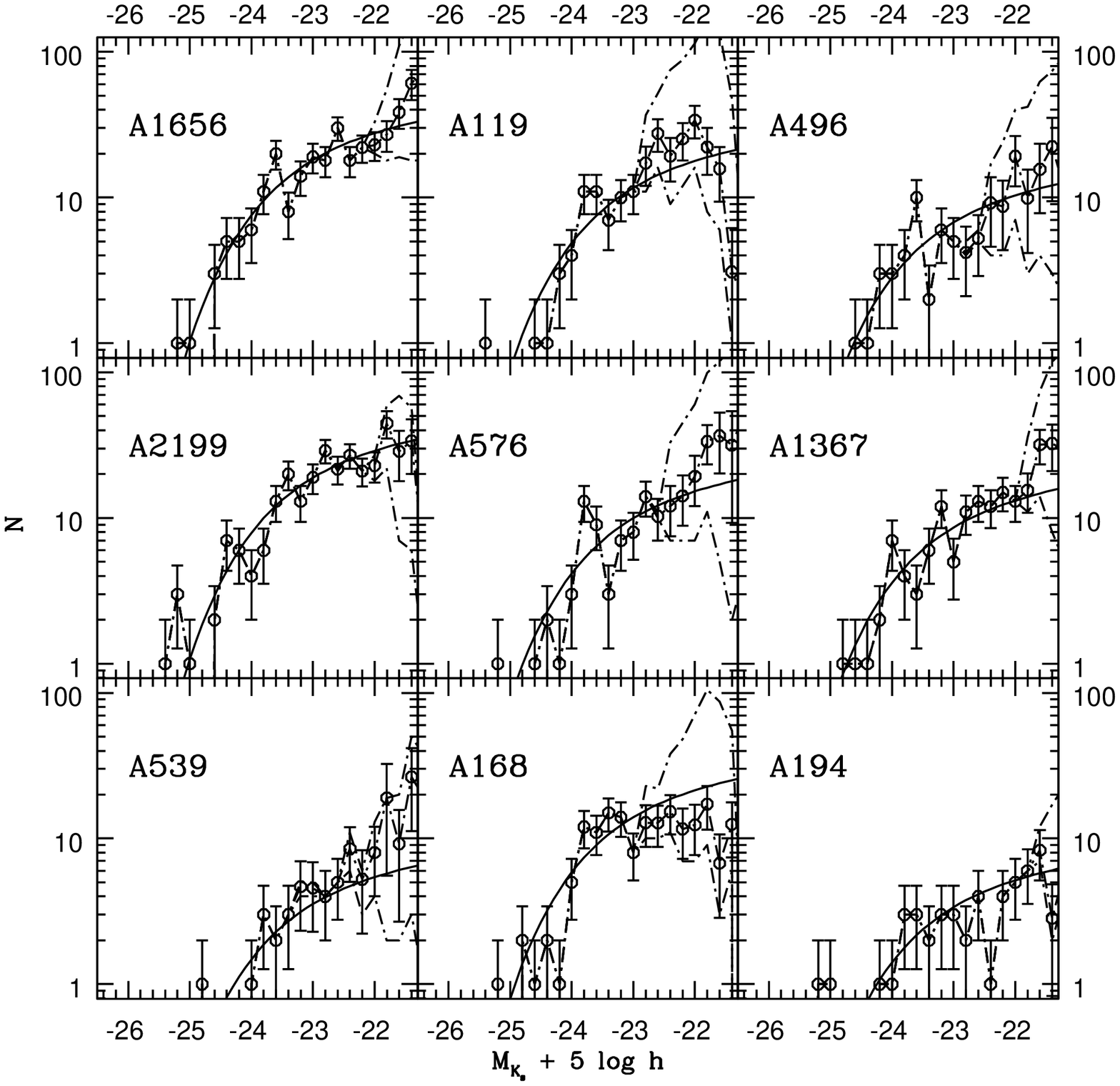,angle=0,width=6.5cm}}
  \caption{$K_s$-band luminosity functions for the CAIRNS clusters.
The left panel shows galaxies in the virial regions (projected within
$r_{200}$); the right panel shows galaxies in the infall regions
(projected outside $r_{200}$).  Solid lines show the shape of the
2dF/2MASS $K_s$-band LF with arbitrary normalization.}
\label{fig:klfncomp}
\end{figure}

When using the mass-to-light ratio in clusters to estimate $\Omega_m$,
one must determine whether the luminosity functions of field and
cluster galaxies differ significantly.  Given the well-known
morphology-density relation, it is possible that the two LFs differ
significantly.  We use 2MASS photometry to determine the near-infrared
luminosity functions for the CAIRNS clusters and infall regions
(\cite{cairnsii}).  The cluster and infall region LFs are very similar
to each other (Figure \ref{fig:klfncomp}) and to the field LF for
relatively bright galaxies ($M_{K_s}\lesssim M_{K_s}^*+2$).  Because
of this similarity, we can correct for the luminosity in faint
galaxies using the field LF.  Our redshift surveys include 60--85\% of
the total light.

\section{Near-Infrared Mass-to-Light Profiles}\label{sec:mlprof}

\begin{figure}
\centerline{\psfig{file=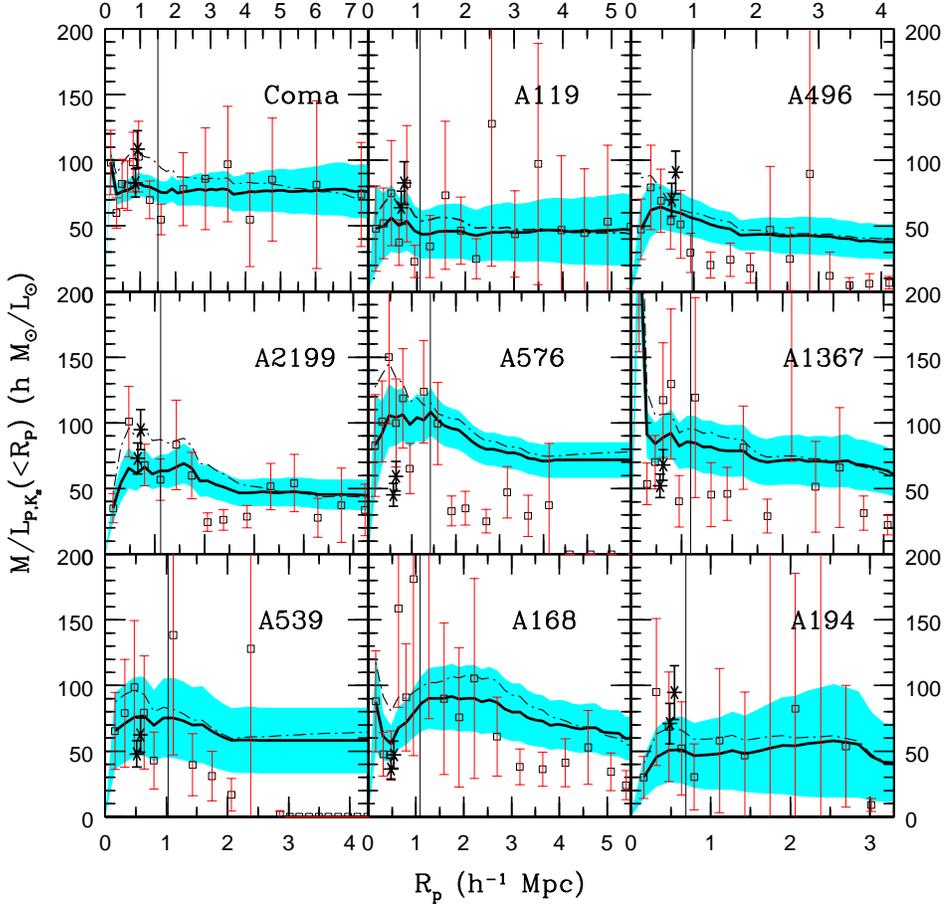,angle=0,width=12.5cm}}
  \caption{Mass-to-light ratio in $K_s$ band as a function of radius for
the CAIRNS clusters. The solid lines show the caustic mass profile
$M(<r)$ divided by the projected luminosity profile $L_{K_s}(<R_p)$
and the shaded regions indicate the associated 1-$\sigma$
uncertainties.  The open squares show the mass-to-light ratio
$M(r,r+dr)/L_{K_s}(R_p,R_p+dR_p)$ in radial shells.  The dash-dotted
line shows the projected best-fit Hernquist mass profile $M_H(<R_p)$
divided by $L_{K_s}(<R_p)$. The solid stars show the mass-to-light
ratio calculated using the X-ray temperature and the mass-temperature
relation to estimate the mass.  The lower point shows
$M_X(<r_{500})/L_{K_s}(<R_{500})$ and the upper point shows
$M_X(<R_{500})/L_{K_s}(<R_{500})$ assuming $M_X(<R_{500}) = 1.3
M_X(<r_{500})$ as is true for an NFW mass profile with $c$=5.}\label{fig:mlkprof}
\end{figure}

Both the surface number density and surface luminosity density
profiles of cluster/infall region members are more extended than the
mass profiles of Paper I.  Figure \ref{fig:mlkprof} shows the
mass-to-light profiles of the CAIRNS clusters.  The mass-to-light
profiles are either flat or show a decreasing $M/L_K$ with increasing
radius.  The mean mass-to-light ratio within $r_{200}$ is a factor of
$1.8\pm0.3$ larger than the mean value outside $r_{200}$.

The decreasing mass-to-light profiles could be caused by gradients in
stellar populations with radius; such effects have been invoked to
account for a similar result for B band mass-to-light profiles in
simulations (\cite{bahcall2000}).  In K band, however, changes in
stellar populations with radius are expected to change the mean
stellar mass-to-light ratio by $\lesssim 20\%$.  Thus, the CAIRNS
mass-to-light profiles provide tentative evidence for variations in
the efficiency of galaxy formation and/or disruption.  Environments
with higher virial temperatures (like cluster centers) are more
efficient at disrupting galaxies and/or less efficient at forming
them.  

A related trend has been noted by Lin et al.~(2003, 2004), who show
that K-band mass-to-light ratios within $r_{500}$ increase with
increasing cluster mass.  This result indicates that more massive
clusters (with larger virial temperatures) have less efficient galaxy
formation and/or more efficient galaxy disruption.  We confirm this
trend in Rines et al.~(2004).  Because cluster infall regions should
be composed of galaxies inhabiting less massive systems
(\cite{rines01b,rines02}) and/or regions with lower virial
temperatures, the above trend predicts that cluster infall regions
should have smaller mass-to-light ratios than virial regions,
consistent with the CAIRNS results.

\begin{figure}
\centerline{\psfig{file=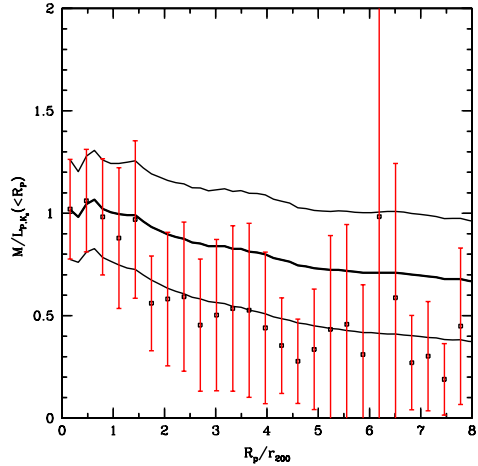,angle=0,width=6.5cm}}
  \caption{Mass-to-light ratio in $K_s$ band as a function of radius for
the combines CAIRNS cluster. The thick and thin solid lines show the caustic
mass profile $M(<r)$ divided by the projected luminosity profile
$L_{K_s}(<R_p)$ and the associated 1-$\sigma$ uncertainties.  The open
squares show the mass-to-light ratio $M(r,r+dr)/L_{K_s}(R_p,R_p+dR_p)$
in radial shells.}
\label{fig:combomlk}
\end{figure}

Figure \ref{fig:combomlk} shows the mass-to-light profile of the
combined CAIRNS cluster (normalized to unity at $r_{200}$).  The
mass-to-light profile clearly decreases with radius.  The combined
cluster should be less susceptible to substructure than the individual
clusters.  Figure \ref{fig:combomlk} shows that the mass-to-light
ratio in shells decreases by about a factor of 2 from the virial
region to the infall region, consistent with the results for
individual clusters.  Taking the mass-to-light ratio in cluster infall
regions as an estimate of the global value and the SDSS luminosity
density extrapolated to $K_s$ band (\cite{blanton03}), we estimate
$\Omega_m=0.10\pm0.02$ (statistical).  This estimate is somewhat lower
if we use the $K_s$ band luminosity density of either Cole et
al.~(2001) or Kochanek et al.~(2001).  We discuss potential systematic
effects in Rines et al.~(2004).  Most of these effects would flatten
the observed mass-to-light profiles relative to the true profiles,
suggesting that the observed decrease in mass-to-light ratio with
radius is real.

\section{Conclusions}\label{sec:conclusions}

Cluster infall regions contain more galaxies than their virial
regions.  If currently popular cosmological models are correct, the
mass in infall regions will eventually accrete onto the parent
clusters, and their final masses will increase by a factor of about 2.

Near-infrared luminosity functions depend only weakly on environment,
at least at the bright end.  We show that near-infrared light is more
extended than mass in cluster infall regions, suggesting environmental
dependence of the efficiency of galaxy formation and/or disruption.
If more efficient galaxy disruption is responsible, intracluster stars
might be a significant component of stellar mass in clusters.  The
mass-to-light ratios in infall regions suggest a low $\Omega_m\sim
0.1$.  Future work is needed to determine the significance of the
conflict of this result with the currently favored $\Omega_m\sim 0.3$.

\begin{acknowledgments}
I would like to thank Margaret Geller and Antonaldo Diaferio for their
many contributions as my primary collaborators in this work.  I would
also like to thank Tom Jarrett, Michael Kurtz, Joe Mohr, Gary Wegner,
and John Huchra for their contributions to the CAIRNS project.  Perry
Berlind, Mike Calkins, and Susan Tokarz collected and reduced most of
the spectroscopic data.  I thank the Smithsonian Institution for
support.
\end{acknowledgments}


\begin{thebibliography}{}

\bibitem[{{Bahcall} {et~al.} 2000}]{bahcall2000}
{Bahcall}, N.~A., {Cen}, R., {Dav{\' e}}, R., {Ostriker}, J.~P., \& {Yu}, Q.
  2000, \textit{Astrop.~J.}, 541, 1

\bibitem[{{Bahcall} {et~al.} 1995}]{bld95}
{Bahcall}, N.~A., {Lubin}, L.~M., \& {Dorman}, V. 1995, \textit{Astrop.~J.}, 447, L81

\bibitem[{{Balogh} {et~al.} 2004}]{balogh03}
{Balogh}, M. {et~al.} 2004, \textit{MNRAS}, 348, 1355

\bibitem[{{Biviano} \& {Girardi} 2003}]{bg03}
{Biviano}, A. \& {Girardi}, M. 2003, \textit{Astrop.~J.}, 585, 205

\bibitem[{{Blanton} {et~al.} 2003}]{blanton03}
{Blanton}, M.~R. {et~al.} 2003, \textit{Astrop.~J.}, 592, 819

\bibitem[{{Carlberg} {et~al.} 1997}]{cye97}
{Carlberg}, R.~G., {Yee}, H.~K.~C., \& {Ellingson}, E. 1997, \textit{Astrop.~J.}, 478, 462

\bibitem[{{Cole} {et~al.} 2001}]{twomdflfn}
{Cole}, S. {et~al.} 2001, \textit{MNRAS}, 326, 255

\bibitem[{{Diaferio} 1999}]{diaferio1999}
{Diaferio}, A. 1999, \textit{MNRAS}, 309, 610

\bibitem[{{Diaferio} \& {Geller} 1997}]{dg97}
{Diaferio}, A. \& {Geller}, M.~J. 1997, \textit{Astrop.~J.}, 481, 633

\bibitem[{{Drinkwater} {et~al.} 2001}]{drink}
{Drinkwater}, M.~J., {Gregg}, M.~D., \& {Colless}, M. 2001, \textit{Astrop.~J.}, 548, L139

\bibitem[{{Ellingson} {et~al.} 2001}]{ellingson01}
{Ellingson}, E., {Lin}, H., {Yee}, H. K.~C., \& {Carlberg}, R.~G. 2001, \textit{Astrop.~J.},
  547, 609

\bibitem[{{Geller} {et~al.} 1999}]{gdk99}
{Geller}, M.~J., {Diaferio}, A., \& {Kurtz}, M.~J. 1999, \textit{Astrop.~J.}, 517, L23

\bibitem[{{G{\' o}mez} {et~al.} 2003}]{gomez03}
{G{\' o}mez}, P.~L. {et~al.} 2003, \textit{Astrop.~J.}, 584, 210

\bibitem[{{Girardi} {et~al.} 2000}]{g2000}
{Girardi}, M., {Borgani}, S., {Giuricin}, G., {Mardirossian}, F., \&
  {Mezzetti}, M. 2000, \textit{Astrop.~J.}, 530, 62

\bibitem[{{Gray} {et~al.} 2004}]{gray04}
{Gray}, M.~E. and {Wolf}, C. and {Meisenheimer}, K. and {Taylor}, A. and 
	{Dye}, S. and {Borch}, A. and {Kleinheinrich}, M.
  2004, \textit{MNRAS}, 347, L73

\bibitem[{{Gunn} \& {Gott} 1972}]{gunngott}
{Gunn}, J.~E. \& {Gott}, J.~R.~I. 1972, \textit{Astrop.~J.}, 176, 1

\bibitem[{{Hernquist} 1990}]{hernquist1990}
{Hernquist}, L. 1990, \textit{Astrop.~J.}, 356, 359

\bibitem[{{Katgert} {et~al.} 2003}]{katgert03}
{Katgert}, P., {Biviano}, A., \& {Mazure}, A. 2003, \textit{Astrop.~J.}, 600, 6457

\bibitem[{{Kauffmann} {et~al.} 1999a}]{kauffmann1999a}
{Kauffmann}, G., {Colberg}, J.~M., {Diaferio}, A., \& {White}, S. D.~M.
  1999a, \textit{MNRAS}, 303, 188

\bibitem[{{Kauffmann} {et~al.} 1999b}]{kauffmann1999b}
---. 1999b, \textit{MNRAS}, 307, 529

\bibitem[{{Kneib} {et~al.} 2003}]{kneib03}
{Kneib}, J.-P. {et~al.} 2003, \textit{Astrop.~J.}, 598, 804

\bibitem[{{Kochanek} {et~al.} 2001}]{twomasslfn}
{Kochanek}, C.~S. {et~al.} 2001, \textit{Astrop.~J.}, 560, 566

\bibitem[{{Kravtsov} \& {Klypin} 1999}]{kk1999}
{Kravtsov}, A.~V. \& {Klypin}, A.~A. 1999, \textit{Astrop.~J.}, 520, 437

\bibitem[{{Lewis} {et~al.} 2002}]{lewis02}
{Lewis}, I. {et~al.} 2002, \textit{MNRAS}, 334, 673

\bibitem[{{Lin} {et~al.} 2003}]{lin03}
{Lin}, Y., {Mohr}, J.~J., \& {Stanford}, S.~A. 2003, \textit{Astrop.~J.}, 591, 749

\bibitem[{{Lin} {et~al.} 2004}]{lin04}
{Lin}, Y., {Mohr}, J.~J., \& {Stanford}, S.~A. 2004, \textit{Astrop.~J.}, in press, astro-ph/0402308

\bibitem[{{Navarro} {et~al.}(1997)}]{nfw97}
{Navarro}, J.~F., {Frenk}, C.~S., \& {White}, S. D.~M. 1997, \textit{Astrop.~J.}, 490, 493

\bibitem[{{Reisenegger} {et~al.} 2000}]{rqcm}
{Reisenegger}, A., {Quintana}, H., {Carrasco}, E.~R., \& {Maze}, J. 2000, \textit{Astron.~J.},
  120, 523

\bibitem[{{Rines} {et~al.} 2003}]{cairnsi}
{Rines}, K., {Geller}, M.~J., {Diaferio}, A., \& {Kurtz}, M.~J. 2003, \textit{Astron.~J.}, 126,
  2152

\bibitem[{{Rines} {et~al.} 2002}]{rines02}
{Rines}, K., {Geller}, M.~J., {Diaferio}, A., {Mahdavi}, A., {Mohr}, J.~J., \&
  {Wegner}, G. 2002, \textit{Astron.~J.}, 124, 1266

\bibitem[{{Rines} {et~al.} 2000}]{rines2000}
{Rines}, K., {Geller}, M.~J., {Diaferio}, A., {Mohr}, J.~J., \& {Wegner}, G.~A.
  2000, \textit{Astron.~J.}, 120, 2338

\bibitem[{{Rines} {et~al.} 2001a}]{rines01a}
{Rines}, K., {Geller}, M.~J., {Kurtz}, M.~J., {Diaferio}, A., {Jarrett}, T.~H.,
  \& {Huchra}, J.~P. 2001a, \textit{Astrop.~J.}, 561, L41

\bibitem[{{Rines} {et~al.}(2001b}]{rines01b}
{Rines}, K., {Mahdavi}, A., {Geller}, M.~J., {Diaferio}, A., {Mohr}, J.~J., \&  {Wegner}, G. 2001b, \textit{Astrop.~J.}, 555, 558

\bibitem[{{Rines} {et~al.} 2004}]{cairnsii}
{Rines}, K. , {Geller}, M.~J., {Diaferio}, A., {Kurtz}, M.~J., \& {Jarrett}, T.~H., 2004, \textit{Astron.~J.}, submitted (astro-ph/0402242)

\bibitem[{{Skrutskie} {et~al.} 1997}]{twomass}
{Skrutskie}, M.~F. {et~al.} 1997, in ASSL Vol. 210: The Impact of Large Scale
  Near-IR Sky Surveys, 25

\bibitem[{{Treu} {et~al.} 2003}]{treu03}
{Treu}, T., {Ellis}, R.~S., {Kneib}, J., {Dressler}, A., {Smail}, I., {Czoske},
  O., {Oemler}, A., \& {Natarajan}, P. 2003, \textit{Astrop.~J.}, 591, 53

\bibitem[{{Tully}  2004}]{tully}
{Tully}, R.~B., 2004, \textit{Astrop.~J.}, submitted, astro-ph/0312441

\bibitem[{{Tustin} {et~al.} 2001}]{tustin}
{Tustin}, A.~W., {Geller}, M.~J., {Kenyon}, S.~J., \& {Diaferio}, A. 2001, \textit{Astron.~J.},
  122, 1289

\bibitem[{{Vedel} \& {Hartwick} 1998}]{vh98}
{Vedel}, H. \& {Hartwick}, F.~D.~A. 1998, \textit{Astrop.~J.}, 501, 509

\bibitem[{{Zwicky} (1933)}]{zwicky1933}
{Zwicky}, F. 1933, Helv.~Phys.~Acta, 6, 110

 
 
\end{thebibliography}
\end{document}